\documentclass[reprint,jap,aip]{revtex4-2} 
\usepackage{amsmath}
\usepackage{amsfonts}
\usepackage{amssymb}  
\usepackage{graphicx}
\usepackage{xcolor}
\usepackage[pdftex,colorlinks=true,linkcolor=blue,citecolor=blue,urlcolor=blue,filecolor=blue,breaklinks=true]{hyperref}  %clickable links in refs
\usepackage{url}
\usepackage{braket}

\usepackage[scientific-notation=true]{siunitx}
\usepackage{setspace}

\usepackage[autostyle, english = american]{csquotes}
\usepackage{mathtools}
\usepackage{cleveref}
\usepackage{siunitx}
\usepackage{float}

\begin{document}

\title{THz range Faraday rotation in the Weyl Semimetal Candidate $\mathrm{Co_2TiGe}$ }

\author{Rishi Bhandia}
\affiliation{The Institute of Quantum Matter, Department of Physics and Astronomy, The Johns Hopkins University, Baltimore, Maryland 21218, USA}

\author{Bing Cheng}
\affiliation{The Institute of Quantum Matter, Department of Physics and Astronomy, The Johns Hopkins University, Baltimore, Maryland 21218, USA}

\author{Tobias Brown-Heft}
\affiliation{Materials Department, University of California-Santa Barbara, Santa Barbara, California 93106, USA}

\author{Shouvik Chatterjee}
\affiliation{Department of Electrical \& Computer Engineering, University of California-Santa Barbara, California 93106, USA}

\author{Christopher J. Palmstr\o m}
\affiliation{Materials Department, University of California-Santa Barbara, Santa Barbara, California 93106, USA}

\author{N. Peter Armitage}
\email{npa@jhu.edu}
\affiliation{The Institute of Quantum Matter, Department of Physics and Astronomy, The Johns Hopkins University, Baltimore, Maryland 21218, USA}

\begin{abstract}
	The $\mathrm{Co_2}$ family of ferromagnetic Heusler alloys have attracted interest due to their fully spin-polarized nature, making them ideal for applications in spintronic devices. More recently, the existence of room temperature time-reversal-breaking Weyl nodes near the Fermi level was predicted and confirmed in these systems. As a result of the presence of these Weyl nodes, these systems possess a non-zero momentum space Berry curvature that can dramatically influence transport properties such as the anomalous Hall effect. One of these candidate compounds is $\mathrm{Co_2 Ti Ge}$. Recently, high quality molecular beam epitaxy-grown thin films of $\mathrm{Co_2 Ti Ge}$ have become available. In this work, we present THz-range measurement of MBE-grown $\mathrm{Co_2 Ti Ge}$ films.  We measure the THz-range Faraday rotation, which can be understood as a measure of the anomalous Hall effect. We supplement this work with electronic band structure calculations showing that the principal contribution to the anomalous Hall effect in the this material stems from the Berry curvature of the material.  Our work shows that this class of Heusler materials shows promise for Weyl semimetal based spintronics.
\end{abstract}

\date{\today}
\maketitle

\section{Introduction}
	
Topological phases of matter have been of great current interest in condensed matter physics. One such manifestation of this non-trivial topological order are Weyl semimetals. These materials arise when the conduction and valence bands of a material touch at a singular point near the Fermi energy in the Brillouin zone~\cite{Armitage2018}. This results in an effective Hamiltonian about this point where the energy has a linear dependence on crystal momentum. These band touchings are called Weyl nodes, and they act as magnetic monopoles (sources and sinks) of an effective magnetic field, the Berry curvature, arising due to the momentum dependence of the Bloch functions in the Brillouin zone~\cite{Armitage2018, Burkov2018}. A requirement for the formation of Weyl nodes in a material is the breaking of inversion or time-reversal symmetry. While inversion breaking Weyl semimetals such as $\mathrm{TaAs}$ have been discovered,~\cite{xu_discovery_2015, lv_experimental_2015} good examples of the theoretically simpler magnetic Weyl semimetals have remained more elusive.

\begin{figure}[ht]
\begin{center}
	\includegraphics[width = 0.48\textwidth]{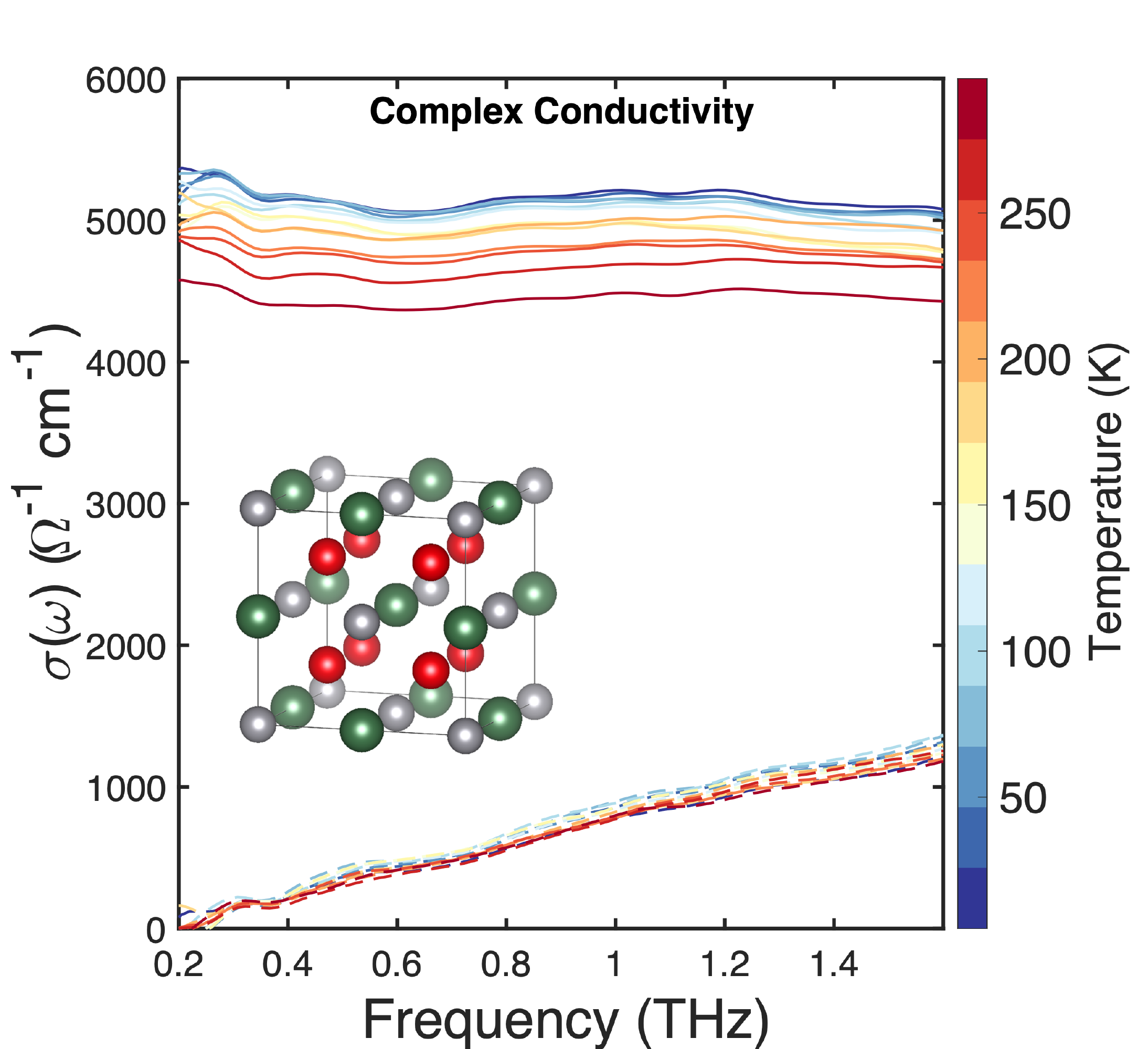}
\end{center}
	 \caption{ Complex conductivity of the $\mathrm{Co_2 TiGe}$ thin film samples   measured using TDTS.  The real part of the conductivity (solid lines) is relatively flat in frequency and exhibits only a weak temperature dependence. The imaginary part of the conductivity (dashed lines) shows a similar temperature dependence. The inset figure shows the $\mathrm{Co_2 Ti Ge}$ Heusler crystal structure, with Co in red, Ti in green, and Ge in grey.}\label{fig:tempConductance}
\end{figure}

The class of Heusler materials, a family of ternary intermetallics, has become a focus of research due to the wide variety of elements that can be incorporated into them.  Due to this flexibility, these materials have been found to exhibit a wide variety of phenomena ranging from half-metallicity~\cite{felser_new_2007} to superconductivity~\cite{winterlik_superconductivity_2009} with potential applications in thermoelectrics~\cite{sakurada_effect_2005} and spintronics~\cite{felser_basics_2015}. They have also drawn interest due to their potential to host non-trivial topological phases because of their band structure's similarity to that of the III-V semiconductors~\cite{chadov_tunable_2010}.  This is due to the fact that the full Heusler with chemical formula $\mathrm{X_2 Y Z}$, forms a zinc-blende lattice with two interpenetrating fcc sublattices as shown in the inset of Fig. \ref{fig:tempConductance}. Thus, Heuslers allow for a great range of properties and tunability while keeping many of the features of III-V semiconductors.

$\mathrm{Co_2}$ Heusler alloys such as $ \mathrm{Co_2TiGe}$ are ferromagnetic with high Curie temperatures~\cite{wurmehl_investigation_2006} and have drawn a great deal of interest due to their half metallic behavior, the property where the minority spin carrier experiences a band gap while the majority spin carrier is gapless~\cite{fecher_slater-pauling_2006, barth_joachim_anomalous_2011, lee_magnetic_2005, kandpal_calculated_2007} Thus, these materials have the potential to be used as the basis for spintronic devices. First principles calculations have shown that these $\mathrm{Co_2}$ alloys could potentially host Weyl fermions that would be accessible at room temperature~\cite{chang_room-temperature_2016, wang_time-reversal-breaking_2016}. More  recently a member of the family, $\mathrm{Co_2 MnGa}$, was confirmed to be a time-reversal-symmetry breaking Weyl semimetal through the existence of "drum-head" surface states~\cite{belopolski_discovery_2019}. Furthermore as these materials are magnetically soft, they also open opportunities to utilize external magnetic field to manipulate the position of Weyl nodes.
 
One potential manifestation of this Weyl physics is the anomalous Hall Effect (AHE)~\cite{burkov_anomalous_2014}.  The AHE occurs when a material exhibits a Hall conductivity, $\sigma_{xy}$ that depends on its magnetization rather than an applied external magnetic field. At a microscopic level, the contributions to the AHE can be separated into three parts:  intrinsic, skew-scattering, and side-jump~\cite{nagaosa_anomalous_2010}.  The former contribution stems from the Berry curvature of the wave functions. Thus, this can be used to probe effects of the Weyl nodes that are theorized to exist within the band structure. However, this analysis is complicated by the fact that it is difficult to distinguish between the intrinsic and side-jump contribution in DC experiments, so AC transport measurements can be an important tool to help distinguish between the different contributions to the AHE. In addition, an investigation of these materials' THz range frequency dynamics is important to understand these material's suitability for usage in next generation spintronic devices.  Optical conductivity may allow for a clearer analysis as by measuring the low-frequency conductivity of a material, it may be possible to distinguish between the intrinsic and side-jump contributions~\cite{nagaosa_anomalous_2010}.

\section{Experimental Details}

\begin{figure}[ht]
\begin{center}
	\includegraphics[width = 0.48\textwidth]{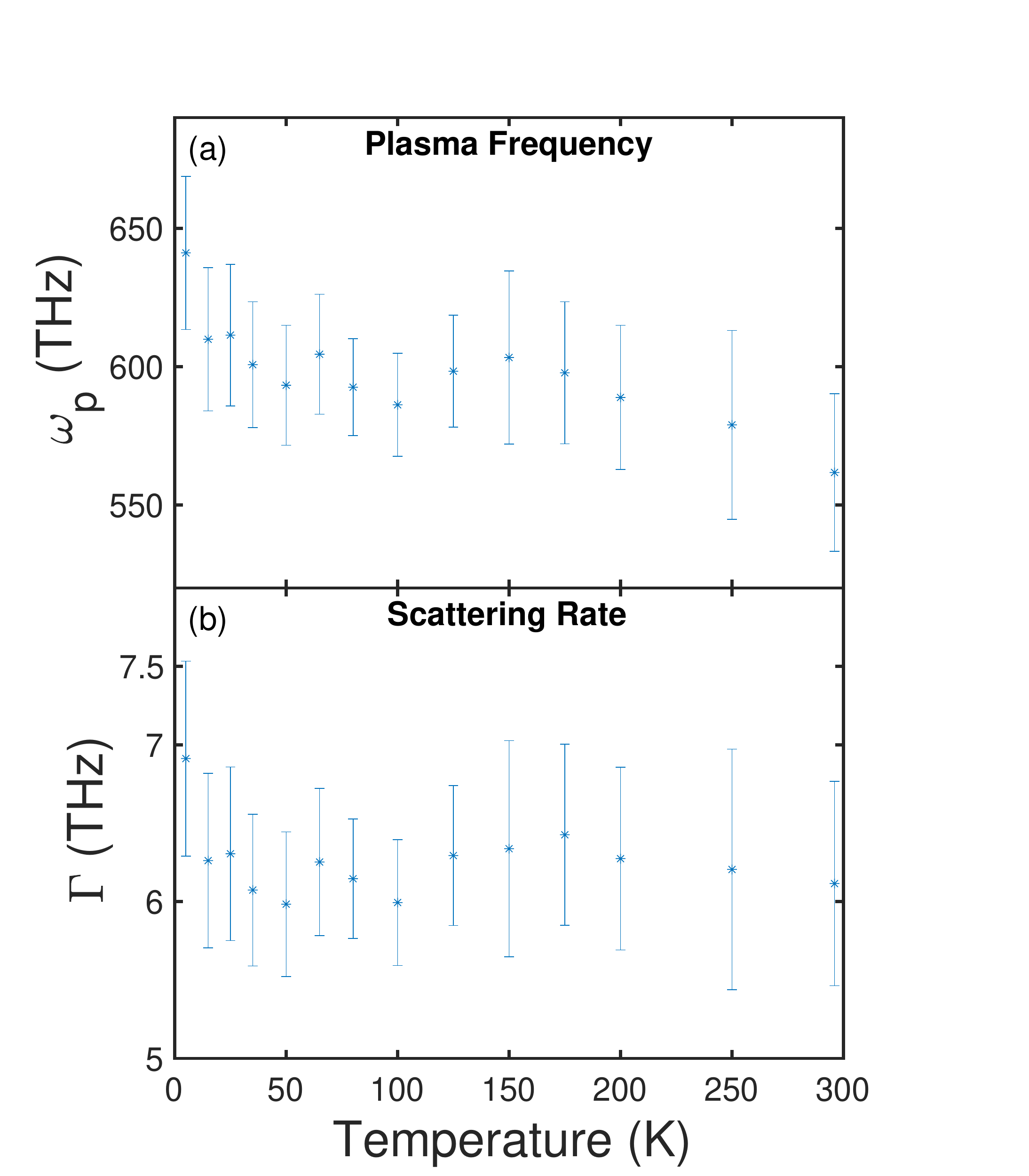}
\end{center}
	 \caption{The fit parameters a single Drude fit for the complex conductivity of the $ \mathrm{Co_2TiGe}$ thin film.  The plasma frequency (a) and the scattering rate (b) both do not show a large dependence on temperature. }\label{fig:drudeParams}
\end{figure}

\begin{figure*}[ht]
\begin{center}
	\includegraphics[width = \textwidth]{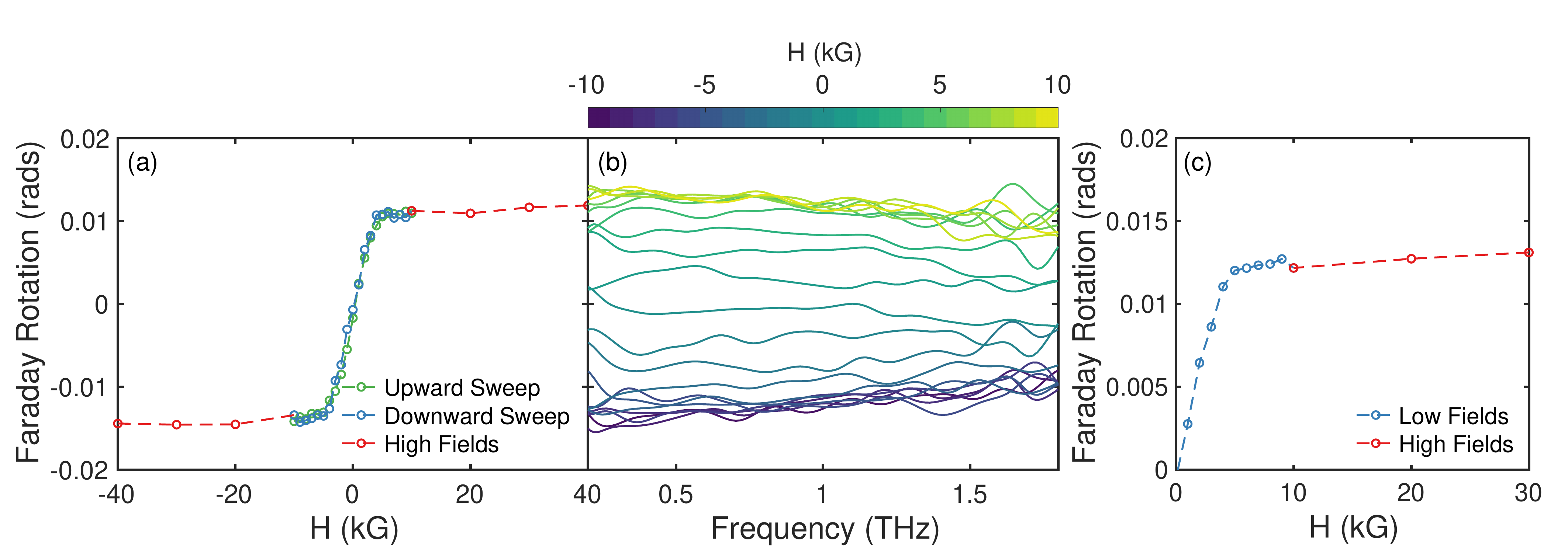}

\end{center}
	 \caption{ The Faraday rotation as measured using THz polarimetry.  (a) The Faraday rotation averaged over the frequency range 0.6 - 1.2 THz as a function of applied field. The saturation at high fields shows that the Faraday rotation is following the magnetization and thus is related to the anomalous Hall effect in the material. (b) The anti-symmetrized Faraday rotation, where opposite fields have been subtracted. We see a clear saturation at fields above $\pm40$ kG.   (c) Averaged anti-symmetrized Faraday rotation, again averaged over the frequency range 0.6 - 1.2 THz.} \label{fig:FaradayRotation}
\end{figure*}

In this work, we present data from time-domain terahertz spectroscopy (TDTS) experiments conducted on high-quality, 72.3 nm thick, thin films of $ \mathrm{Co_2TiGe}$ grown by molecular beam epitaxy (MBE) on a substrate of MgO (001)\cite{logan_growth_2017}. These experiments were conducted with a home-made time-domain terahertz spectrometer with a  7 T magnet in closed cycle cryostat.\cite{morris_polarization_2012} By measuring the transmitted electric field through the sample and an appropriate reference substrate, we are able to calculate the complex transmission coefficient. From this, then the complex conductivity was calculated using appropriate equation in the thin film limit.  In zero magnetic field we use the standard expression $T\left(\omega\right) = \frac{1 + n}{1 + n +Z_0 d \sigma\left(\omega\right)} \mathrm{exp}\left[ i\frac{\omega }{c} \left( n - 1 \right) \Delta L\right]$. Here, $T\left(\omega\right)$ is the complex transmission referenced to the substrate, $n$ is the index of the refraction of the substrate, $d$ is the thickness of the thin film, $Z_0$ is the impedance of the free space with a value of roughly $377\  \Omega$, and $\Delta L$ is a small thickness difference between the sample substrate and reference substrate.

 To examine the THz range $\sigma_{xy}(\omega)$ of the material we measured the Faraday rotation of the THz light transmitted through our sample under finite magnetic field. The magnetic field was applied parallel to the direction of propagation of light in the Faraday geometry. As the Faraday rotation is the angle that a linearly polarized wave of light is rotated due to a magnetic field parallel to its direction of propagation, it is directly related to the transverse conductivity. In this geometry, utilizing a rotating wire-grid polarizer to modulate the polarization, we are then able to measure the $E_{x}$ and the $E_{y}$ components of the transmitted electric fields~\cite{morris_polarization_2012}.  By symmetry considerations of the material, we can constrain the Jones transmission matrix to have only two free parameters, $T_{yx}$ and $T_{xx}$ and calculate these matrix elements~\cite{armitage2014constraints}. Then, to extract the Faraday rotation from the measured complex transmission we used the expression $\theta_F = \mathrm{Re}\left[\mathrm{arctan}\left(T_{yx} / T_{xx}\right)\right]$.
	
	  % For a conducting, optically-thin film grown on a substrate such as the films used in this study, one can use the Tinkham equation to exactly determine the the complex optical conductivity or conductance of a material. As we compare the transmitted electric fields through the sample with that of a bare substrate, it becomes important that the thickness difference between the sample and reference is precisely known so that we can determine the phase of the complex conductivity.  In this case, we determined the difference in thickness between the sample and reference substrate by examining the difference in time between initial terahertz pulse and the pulse that is reflected at the substrate-air interface. As shown in \Cref{fig:ReflectionTimeTraces}, we measure the difference time between the original terahertz pulse and the reflected pulse. From this, the thickness of the substrate can be computed if its index of refraction is known. As these thin films were grown on a standard substrate, MgO whose index of refraction is known in the THz frequency range, this allowed us to determine the thickness difference between the sample and reference substrate.
	  
Fig. \ref{fig:tempConductance} depicts the real and imaginary parts of the optical conductivity over the temperature range 5 K - 300 K. The overall magnitude of the conductivity increases with decreasing temperature. The real part of the conductivity is flat and the imaginary part is small, consistent with the scattering rate for electrons well above the measured spectral region. The conductivity can then be fit with the Drude model for a free electron gas
  \begin{align}
 	 \sigma(\omega) = \epsilon_0  \left[\frac{-\omega_p^2}{i\omega - \Gamma}   - i(\epsilon_\infty - 1)\omega\right],
  \end{align} 
where $\omega_p$ is the plasma frequency, $\Gamma$ is the scattering rate, and $\epsilon_\infty$ is the high frequency dielectric constant which accounts for the effects of higher band interband transitions on the low frequency dielectric constant.  The results of these fits are shown in Fig. \ref{fig:drudeParams}.  As we can see, neither plasma frequency nor scattering rate show a significant dependence on temperature.  The plasma frequency shows about a 10$\%$ enhancement as temperature decreases that is consistent with the overall increase in conductivity as temperature decreases.  The errors of the fit parameters are derived from the 95\% confidence intervals derived from the fitting routine.

\begin{figure}[ht]
\begin{center}
	\includegraphics[width = 0.48\textwidth]{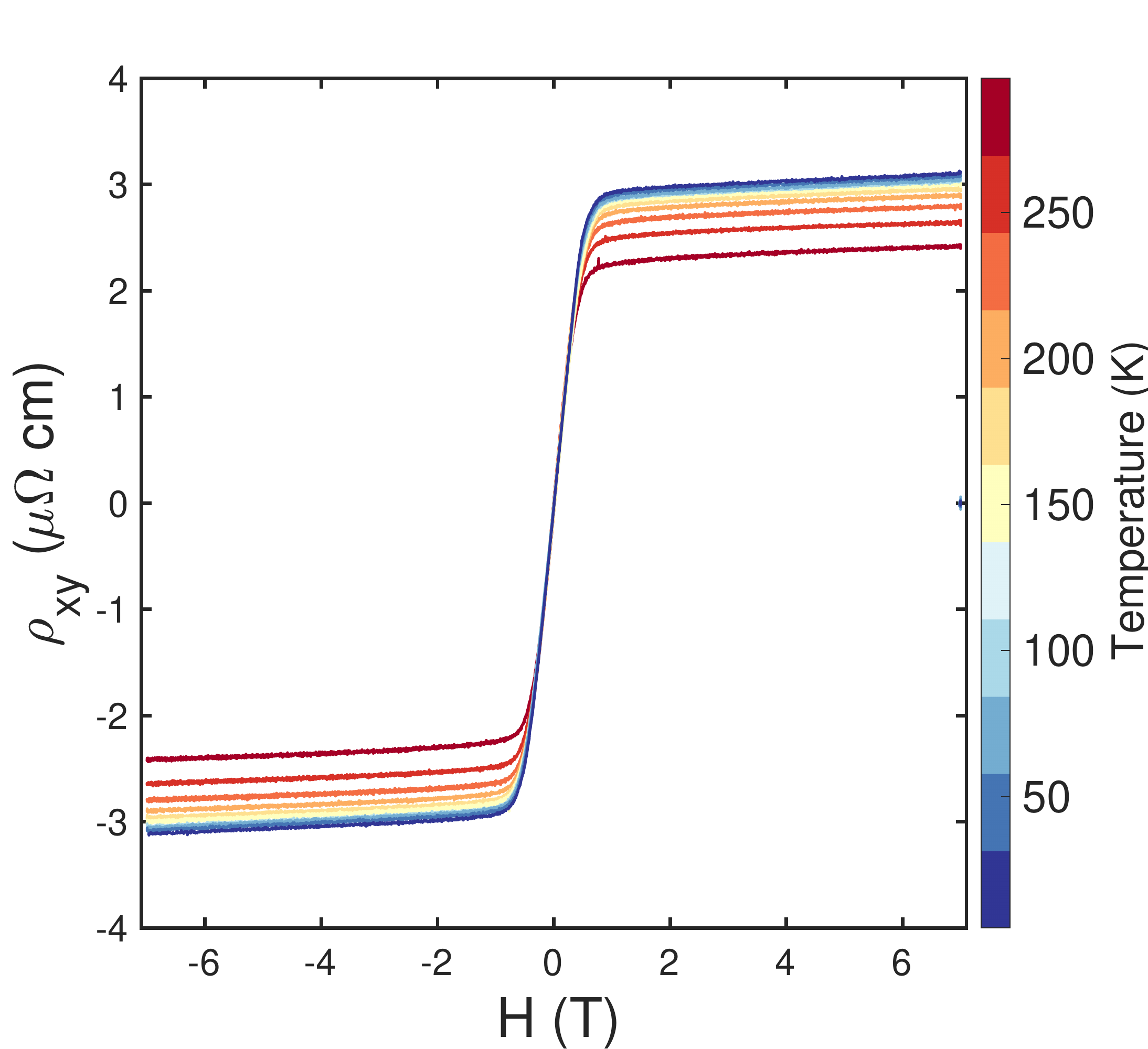}
\end{center}
	 \caption{DC anomalous Hall effect data measured in Hall bar geometry with magnetic field in [001] direction. }\label{fig:HallPlot}
\end{figure}

\begin{figure*}[ht]
\begin{center}
	\includegraphics[width = \textwidth]{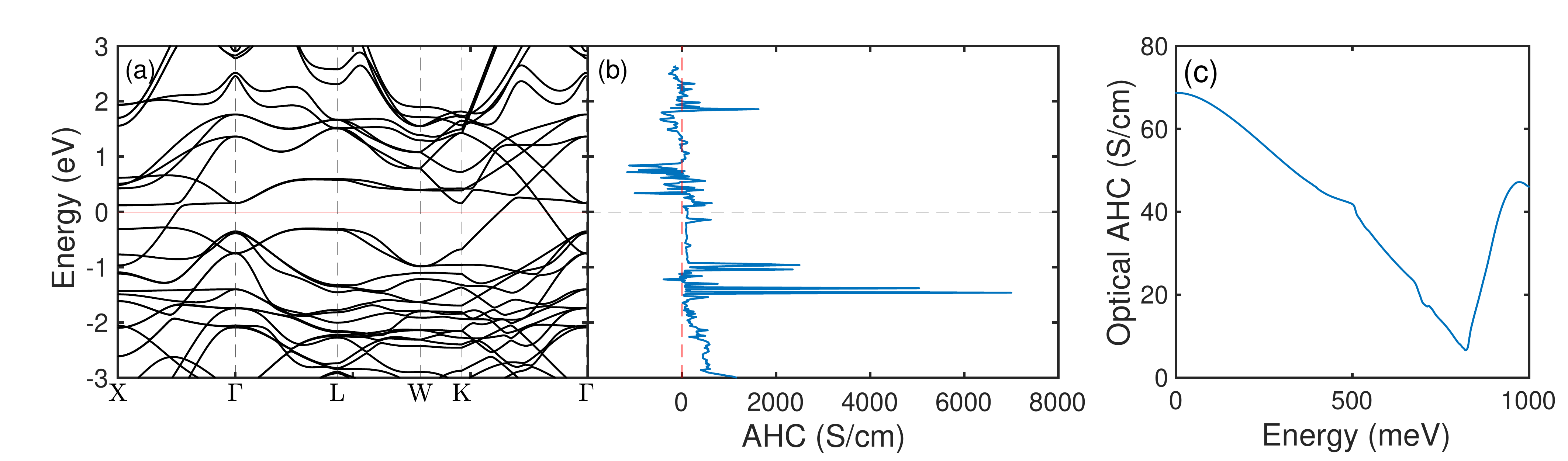}
	\caption{The results of the band structure calculations presented in plots. (a) The first principles calculated band structure along symmetry directions with magnetization in the (001) direction and spin orbit coupling turned on. Spin orbit coupling was as implemented in Quantum Espresso. The zero level is referenced to the Fermi energy calculated using Quantum Espresso.(b) The calculated intrinsic DC anomalous Hall effect as a function of Fermi energy calculated using Wannier90. (c) The optical intrinsic anomalous Hall conductivity calculated using Wannier90.}\label{fig:dftPlots}
\end{center}
\end{figure*}

In Fig. \ref{fig:FaradayRotation}, we show the results of our polarimetry experiments. Fig. \ref{fig:FaradayRotation} (a) gives the Faraday rotation measured with field being swept either upward or downward.  We see no sign of a hysteresis as expected in a soft ferromagnetic thin film due to its in-plane easy axis, consistent with DC Hall effect data shown in Fig. \ref{fig:HallPlot}. This is consistent with easy axis of these films lying in-plane, in the [110] direction as reported earlier \cite{logan_growth_2017}. Starting from high fields, as the applied magnetic field decreases in magnitude, the magnetization relaxes to an in-plane direction, resulting in almost zero measured Faraday rotation due to the in-plane magnetization.   Fig. \ref{fig:FaradayRotation} (b) shows the Faraday rotation anti-symmetrized, where the up and down sweeps have been subtracted at opposite fields i.e. $ \theta_{\textrm{AS}} \left(H\right) = \left(\theta_{\textrm{up}} \left(H\right) - \theta_{\textrm{down}} \left( -H\right) \right) / 2$ where $\theta$ is the Faraday rotation and $H$ is the magnetic field. Fig. \ref{fig:FaradayRotation} (c) then shows the anti-symmetrized Faraday rotation averaged over the frequency range 0.6 - 1.2 THz, again showing the saturation that we saw earlier at fields above  5 kG. The Faraday rotation's saturation at low fields is consistent with the expectation that the AHE depends on the magnetization of the material. As the Faraday rotation can be also described by the expression, 
\begin{align}
 \theta_F \approx \frac{d\sigma_{xy} Z_0}{1 + n_{sub} + d\sigma_{xx} Z_0}, \label{eq:FaradayThinFilm}
 \end{align}
  we can see that in the zero frequency limit, it describes the Hall angle (if one neglects effects due the substrate).
    
To provide complementary information to our THz measurements, we used the open-source package \textit{Quantum Espresso}~\cite{giannozzi_quantum_2020, giannozzi_advanced_2017,giannozzi_quantum_2009} to calculate electronic band-structure calculations for $ \mathrm{Co_2TiGe}$ under the density functional theory (DFT) framework. The projector-augmented wave method (PAW) was used with fully relativistic pseudo potentials, spin orbit coupling, and the exchange functional was approximated using the generalized gradient approximation (GGA). To the compute transport properties, we used the code Wannier90~\cite{pizzi_wannier90_2020} to project the calculated wavefunctions onto maximally localized Wannier functions. From this, a $50 \times 50 \times 50$ adaptive grid was used to sample the Berry connection within the Brillouin zone.  We then computed the anomalous Hall conductivity numerically using Wannier90 with the Kubo-Greenwood formula as follows:

\begin{align}
 	\sigma^{AHC}_{xy}\left(\hbar \omega\right) =  &\frac{i e^2 } {\hbar \Omega_c N_k } \sum_{\textbf{k}, n, m} \frac{  \left( f_{n\textbf{k}} -  f_{m\textbf{k}} \right) \left(\varepsilon_{m \textbf{k}} - \varepsilon_{n \textbf{k}} \right)} {\varepsilon_{m \textbf{k}} - \varepsilon{n \textbf{k}}  - \left(\hbar \omega + i \eta\right)} \nonumber \\ 
&	\times A_{nm, x}(\textbf{k}) A_{mn, y}(\textbf{k}) \label{eq:AHC}
\end{align}
 where $\Omega_c$ is the cell volume, $N_c$ is the number of $\textbf{k}$ points used to sample the Brillouin zone, $f_{n\textbf{k}}$ is the Fermi-Dirac distribution function for the band indexed by $n$, $\eta$ is a smearing parameter, and $A_{mn}(\textbf{k}) = \braket{u_{n\textbf{k}} | i \nabla_{\textbf{k}} | u_{m\textbf{k}} }$ is the matrix generalization of the Berry connection.  For the DC limit as shown in Fig. \ref{fig:dftPlots} (b), we simply take the parameters $\omega = 0$ and $\eta = 0$ in Eq. \ref{eq:AHC}.  In Fig. \ref{fig:dftPlots}a one can see a large complement of bands.  Weyl nodes are located along the $\Gamma - X$ and $\Gamma - K$ directions.   
   
\section{Discussion}
\begin{figure}[h]
\begin{center}
	\includegraphics[width = 0.48\textwidth]{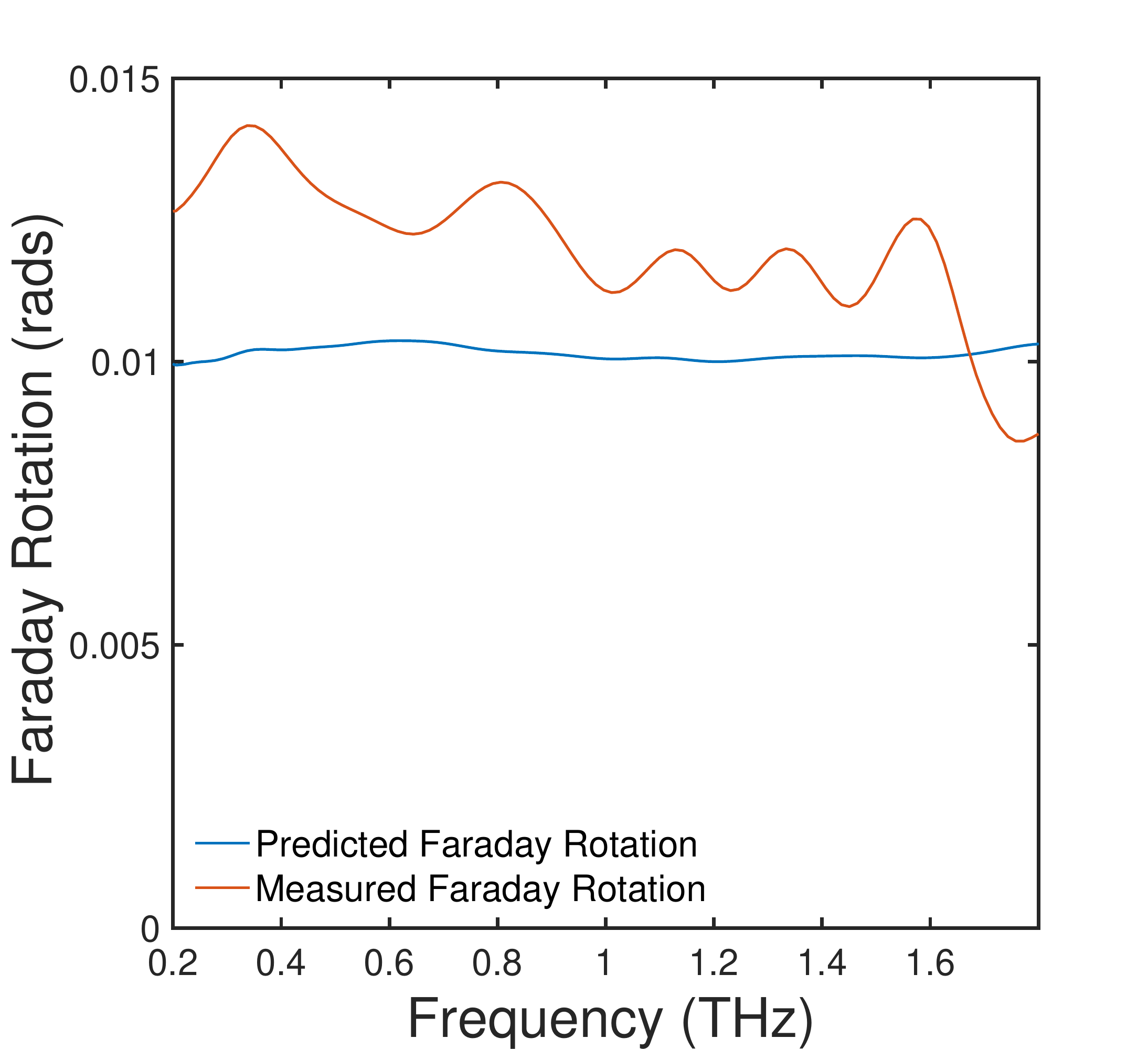}
\end{center}
	 \caption{Predicted Faraday rotation using calculated intrinsic anomalous Hall effect and measured optical conductivity. The frequency dependent structures in the experimental data reflects systematic errors in the measurement that are amplified due the small overall value of the Faraday rotation.}\label{fig:PredictedFaraday}
\end{figure}

The behavior of these $\mathrm{Co_2 Ti Ge}$ thin films is consistent with the expected behavior for a magnetic metal.  Through the measurement of the Faraday rotation, we understand the low-frequency response of the Hall effect.  Above a low applied field, the Faraday rotation saturates, showing that $\sigma_{xy}$ depends on the magnetization. This suggests that the Faraday rotation is sensitive to the AHE and thus sensitive to the Berry curvature of the material.
		
In the simplest case, for a pair of a Weyl nodes near the Fermi level, the AHE contribution produced by a single pair of Weyl nodes due to the Berry curvature is given by 
\begin{align}
\sigma_{xy} = \frac{e^2 K}{2 \pi h},
\end{align}
 where $K$ is the wavevector between the Weyl nodes in the Brillouin zone.  Using the Faraday rotation we can  estimate the Hall conductivity using the expression provided previously and making the assumption that magnetic field does not affect the magnitude of the longitudinal transport.  We estimate $\sigma_{xy}$  as \SI{83.8} {\ohm^{-1} \cm^{-1}}.  For a single pair of Weyl nodes, this translates to a spacing of $0.25 \frac{\pi}{a}$ where $a$, the lattice constant, is $ \SI{5.830}{\angstrom}$. More generally this can be seen as a measure of the total effective Berry dipole moment, since Weyl nodes are monopoles of Berry curvature and the AHE contributions are additive. However, in $\mathrm{Co_2 TiGe}$ there are believed to be nine pairs of Weyl nodes~\cite{chang_room-temperature_2016} oriented such that some of their contributions partially cancel.  In the upper half of the Brillouin zone, there is predicted to be one Chern charge 2 Weyl node displaced along the $k_z$ direction and two Chern charge 1 Weyl nodes with opposite relative charge in the $k_x-k_z$ plane.   Other Weyl nodes are related to these by the mirror symmetry in the $k_z=0$ plane and $C_4$ giving 9 pairs total.  Using the Weyl node positions calculated via DFT ~\cite{chang_room-temperature_2016}, we estimate a total Berry dipole moment of $1.12 \frac{\pi}{a}$, which is of order the observed value.   Discrepancies with the actual value may arise from the fact that it comes from a difference of large numbers and will have a proportionally larger uncertainty or the fact that some of the nodes (particularly the charge 2 pair), may not give their full contribution to the AHE as they are far from E$_F$.

We can also compare this estimation of $\sigma_{xy}$ with the measured DC values of $\rho_{xy}$. Here, using an average over a frequency range of 0.6 to 1.2 THz of the measured optical conductivity to approximate the zero frequency $\sigma_{xx}$, we obtain an estimated hall resistivity of \SI{3.15} {\micro \ohm \cm}.  This gives a value very close to the measured hall resistivity measured in the hall bar geometry as depicted in Fig \ref{fig:HallPlot}.
 
 Using the band structure calculations, we can make a more detailed analysis of the contributions to the anomalous Hall effect. The expected Faraday rotation can be estimated using the measured optical conductivity and calculated intrinsic optical anomalous Hall conductivity in conjunction with Eq. \ref{eq:FaradayThinFilm}. The results are plotted in Fig. \ref{fig:PredictedFaraday}.  The measured Faraday rotation shows a consistent offset with respect to the calculated Faraday rotation. We believe the small mismatch is due to the contribution of skew-scattering and side-jump. However, it appears that the bulk of the contribution to the AHE in this material stems from the intrinsic Berry curvature of this material.
	
\section{Conclusion}
		We have measured the THz-range optical conductivity of the $\mathrm{Co_2}$ Heusler $\mathrm{Co_2 TiGe}$ using time-domain THz spectroscopy.  These measurements were made possible by MBE-grown thin films of this system becoming available.  We have also measured the THz-range Faraday rotation of this material.  Using DFT calculations, we have demonstrated that the majority contribution to the Faraday rotation stems from the intrinsic Berry phase mechanism for the anomalous Hall effect of the material.  
\section*{Acknowledgements}
	    We would like to thank T. McQueen for very helpful discussions regarding the DFT calculations. Work at JHU was supported as part of the Institute for Quantum Matter, an EFRC funded by the DOE BES under DE-SC0019331.  Work at UCSB was supported by DOE BES under DE-SC0014388. We also acknowledge the use of shared facilities within the National Science Foundation (NSF) Materials Research and Science and Engineering Center (MRSEC) at the University of California: Santa Barbara, DMR 1720256 and the LeRoy Eyring Center for Solid State Science at Arizona State University. 
	    
\section*{Data Availability}
The data that support the findings of this study are available from the corresponding author upon reasonable request.
\bibliography{Co2TiGeFaradayRotation}

\end{document}